\documentclass[superscriptaddress,twocolumn,showpacs,amsmath,amsfonts]{revtex4}
\usepackage{latexsym}

\newtheorem{theorem}{Theorem}
\newcommand{\qed}{\rule{7pt}{7pt}}
\newcommand{\ignore}[1]{}
\newenvironment{proof}{\noindent{\bf Proof}\hspace*{1ex}}{\qed\bigskip}

\def\cP{{\cal P}}

\newcommand{\eps}{\varepsilon}

 \newcommand{\R}{\mathbb{R}}

\renewcommand{\Pr}[1]{\mathbb{P}\left(#1\right)}

\def\be{\begin{equation}}
\def\ee{\end{equation}}
\def\bea{\begin{eqnarray}}
\def\eea{\end{eqnarray}}
\newcommand{\bra}[1]{\mbox{$\left\langle #1 \right|$}}
\newcommand{\ket}[1]{\mbox{$\left| #1 \right\rangle$}}

\begin{document}

\author{Aram Harrow}
\email{aram@mit.edu}
\affiliation{Department of Physics, MIT, 77 Massachusetts Ave., Cambridge, MA 02139, USA \\
\& Department of Computer Science, University of Bristol, Bristol BS8 1TW, UK}
\author{Roberto Oliveira}
\email{rob.oliv@gmail.com}
\affiliation{IBM Watson Research Center, P.O. Box 218, Yorktown
Heights, NY 10598, USA}
% \affiliation{Courant Institute of Mathematical Sciences, Mercer
% Street, New York City, NY  ,USA}

\author{Barbara M. Terhal}
\email{terhal@watson.ibm.com}
\affiliation{IBM Watson Research Center, P.O. Box 218, Yorktown
Heights, NY 10598, USA}

\title{The cryptographic power of misaligned reference frames}

\date{\today}

\begin{abstract}
Suppose that Alice and Bob define their coordinate axes
differently, and the change of reference frame between them is
given by a probability distribution $\mu$ over SO(3).  We show
that this uncertainty of reference frame is of no use for bit
commitment when $\mu$ is uniformly distributed over a (sub)group of
SO(3), but other choices of $\mu$ can give rise to a partially or even
asymptotically secure bit commitment.
\end{abstract}


\maketitle

It has been one of the goals of quantum information theory to find
new cryptographic applications of quantum physics. The prime and
most successful example of such an application is the protocol of
quantum key distribution developed by Bennett \& Brassard in 1984
\cite{BB:84}, which implements a cryptographic primitive that is
impossible to obtain via classical means.

Another cryptographic primitive of great interest is secure bit commitment.
\ignore{In bit commitment protocols, a party called Alice commits
herself to the contents of a message that she will later reveal to
another party, Bob. }Bit commitment protocols typically involve
two phases: a commit phase, in which Alice commits to a bit $b$
(called the message), and the reveal phase, when Bob learns $b$.
The requirements for secure bit commitment are that the protocol
is (at least approximately) {\em sound}, meaning that when both
parties are honest Bob accepts the message $b$; {\em binding},
meaning that after the commit phase, a cheating Alice will never
be able to reliably convince Bob of more than a single value of $b$
(although she may force Bob to abort the protocol); and {\em
concealing}, meaning that a cheating Bob cannot learn the value of
$b$ before the reveal stage, irrespective of his cheating strategy
(of course, no guarantees are possible if both parties cheat).

Unlike the case of key distribution, neither classical nor quantum
resources suffice for secure bit commitment under general
circumstances \cite{mayers:qbcsecurity,LC:qbcsecurity,LC:qbcsecurity2}, not even
when both parties participating in the protocol are restricted by
local superselection rules \cite{KMP:ss}. However, it {\it is}
possible to build secure bit commitment using additional
assumptions, such as the hardness of performing certain
calculations in polynomial-time \cite{blum:bc,naor:bc} or, in the
classical case, the availability of a known, noisy (i.i.d.) channel
between the parties (see Ref. \cite{WNI:bc} and references therein).\\

\indent This Letter is concerned with the possibility of secure quantum
bit commitment when the spatial frames of reference between the parties
are misaligned. Depending on a single qubit's physical
implementation, its definitions of $\ket{0}$ and $\ket{1}$ are
given by a local measurement setting which may be directionally
dependent. For example, a spin qubit realization will depend on
the orientation of locally applied magnetic fields and qubits
based on polarization degrees of freedom will depend on the local
settings of polarization filters. This means that between different locations
there may be a misalignment of such spatial frames. Misaligned reference frames
could also arise in a (special) relativistic setting \cite{PT:rel}. Assume that
Alice's inertial frame moves at relative velocity $\vec{v}$ with respect to Bob and Alice and Bob
have partial information about this velocity. This implies that quantum information sent between
the parties is subject to unitary transformations representing boosts in the Lorentz group.

Several recent papers have studied the problem of misaligned reference frames. Researchers have
considered the task of communicating a reference frame or direction \cite{PS:dir,RG:cc,CPS:refframe,vanenk:refbit},
communication in the presence of misaligned reference frames \cite{BRS:comm},
or the cryptographic use of a private shared reference frame \cite{BRS:cryp}.

In our set-up we assume that Alice and Bob define
their computational bases $\{\ket{0},\ket{1}\}$ differently. That is,
suppose Alice uses $\{\ket{0_A},\ket{1_A}\}$ and Bob uses
$\{\ket{0_B},\ket{1_B}\}$. Let $U=\ket{0_B}\bra{0_A} +
\ket{1_B}\bra{1_A}$ be the $2\times 2$ unitary matrix relating
these bases. If the reference frames are related by a
3-dimensional rotation $R$ taken from probability measure $\mu$,
then via the homomorphism between SU(2) and SO(3), $U \in {\rm
SU(2)}$ represents a sample from $\mu$. The distribution $\mu$ captures
Alice's and Bob's information, which we assume to be the same,
about the misalignment between their reference frames. For example, if Alice and Bob
have no knowledge about their misalignment, they assume that $U$ is uniformly
at random from ${\rm SU}(2)$. Partial but identical knowledge by both parties
can be represented by a more involved distribution $\mu$.
The effect of the misalignment is that everything Alice sends to Bob is
multiplied by $U$ (in his frame of reference) and everything Bob sends to Alice will be
multiplied by $U^\dag$.
This is not unlike a noisy quantum channel. Notice, however, that
unlike an ordinary quantum channel, no quantum information is
destroyed by channel uses. For example, if Bob sends the state
back to Alice, she will recover her original message; or if
Bob later learns an approximate description of $U$, then he can
apply a unitary close to $U^\dag$ and obtain a state close to what
Alice sent. Moreover, if Alice sends $k$ qubits, the `channel'
applies a random $U^{\otimes k}$ to the $k$ qubits, unlike $k$
independent copies of a standard channel. \\

\indent Our main goal in this Letter is to determine whether the no-go
results for bit commitment still hold under the assumption of
misaligned reference frames. Our results are two-fold. On the one
hand, we prove that if $\mu$ is uniform over SU(2) or a subgroup
of SU(2), then the standard security results of bit commitment still hold. On the
other hand, we also show that there exists distributions $\mu$, albeit
somewhat artificial, that give rise to asymptotically secure bit
commitment.

% \section{Uniform rotations do not help}

Let us now prove our first result, namely if the rotation relating
Alice to Bob's reference frame is completely unknown, i.e. $\mu$
is a uniform distribution over (a subgroup of) SU(2), then bit
commitment is impossible.

\begin{theorem}
Let $G$ be a subgroup of {\rm SU(2)}, and $\mu$ the uniform distribution
over $G$. If there exists a bit commitment protocol $\cP$ using
misalignment characterized by $\mu$, then
there exists a protocol $\cP'$ using a standard noiseless quantum
channel and locally aligned reference frames with identical
security parameters.
%parameters $\delta_s$, $\delta_b$ and $\delta_h$.
\label{theo1}
\end{theorem}

The theorem shows that there is no gain in bit commitment security
over what is possible in the standard scenario when the misalignment is
taken uniformly at random in some group, for example rotations around some fixed axis
in the Bloch sphere. Thus, as in the standard case,
perfectly secure bit commitment is impossible for such misalignments
Here is the proof of the theorem:\\

\begin{proof}
Given $\cP$, we construct $\cP'$ as follows.  Alice and Bob choose
random rotations $U_A$ and $U_B$ uniformly at random from $G$.
The protocol $\cP'$ follows $\cP$, but with the following changes:

\begin{itemize}
\item Before sending any qubit, Alice applies $U_A$.
\item After receiving any qubit, Alice applies $U_A^\dag$.
\item Before sending any qubit, Bob applies $U_B$.
\item After receiving any qubit, Bob applies $U_B^\dag$.
\end{itemize}

If both parties are honest, then $\cP'$ functions the same way as
$\cP$, except with the reference frames of the two parties related
by $U_B^\dag U_A$.  Since $U_A$ and $U_B$ are chosen uniformly at
random, $U_B^\dag U_A$ is as well, and the protocol simulates
$\cP$ exactly. Therefore $\cP'$ is just as sound as $\cP$.

Now suppose one party, say Bob, cheats.  We can make no
assumptions about Bob's actions, but since Alice is still honest,
her random selection of $U_A$ is enough to randomize her reference
frame from Bob's point of view.  Furthermore, she otherwise
follows $\cP$, so Bob cannot distinguish Alice's half of $\cP'$
from $\cP$ being executed over a channel with a genuinely random
rotation.  Therefore $\cP'$ is just as concealing as $\cP$.

If instead Alice cheats, the same argument shows that $\cP'$ is
just as binding as $\cP$.
% As always, if both parties cheat, then they deserve whatever they
% get.
\end{proof}

We will now show that the conclusion of Theorem 1 does {\em not}
hold for general distributions $\mu$. Since noisy classical
channels can have a bit commitment capacity \cite{WNI:bc} (which
is achieved by coding), we may expect that a single use of such a
channel could lead to a bit commitment that goes beyond what is
possible without the noisy channel. In fact we will prove a
stronger result, namely that there exists a distribution that allows for
an asymptotically secure bit commitment.

The classical protocols developed in \cite{WNI:bc} are of the
following form, which we shall also employ: Alice chooses a set of
codewords $C_{a,b}$, where $b$ is the message (usually a single
bit) to be committed and $a$ is another index of arbitrary size
that is picked randomly.  She sends $C_{a,b}$ through the channel
to commit to $b$, and reveals $b$ by sending $(a,b)$.  Bob will
typically accept if there is a high likelihood that the noisy
message that he received in the commit phase originated from the
codeword $C_{a,b}$ for the values of $a,b$ that Alice gives him in
the reveal phase. We cannot apply such a protocol directly since
our `misalignment channel' is not i.i.d. (independent and
identically distributed) which rules out asymptotic coding
results. Furthermore, we are considering quantum instead of
classical information.

The latter issue is easily fixed. We will consider the limit in
which the quantum information that Alice sends behaves
classically, that is, Bob can perfectly (in the classical limit)
distinguish the states $\ket{C_{a,b}}$. In case we deal with
`misalignment channels' for photon polarization, this means that
Alice will send Bob a very bright beam of identically polarized
photons, so that Bob can determine this polarization vector with
high accuracy. If we are working with spins, the codewords that we
will use will be spin $J$ particles for $J \rightarrow \infty$.

Since any classical protocol has an underlying quantum mechanical implementation, one cannot
exclude a priori the possibility for a coherent quantum attack, even though this kind of
attack may be technically hard to implement in the classical limit. This kind of attack
has been the main limitation to the security of quantum bit commitment \cite{mayers:qbcsecurity,LC:qbcsecurity,LC:qbcsecurity2}.
In the coherent quantum attack Alice has an additional ancilla entangled with the codeword that
she will send to Bob, i.e. she holds a purification of the state she sends to Bob.
This may enable her to cheat at the revealing stage. However, for the schemes that we will consider, we can easily argue
that such a cheating strategy will not give Alice extra power. We do this by demanding that an honest Bob always
first measures the quantum state he obtains from Alice. Since we are in the classical limit, his measurement (of polarization
or spin-direction) will hardly disturb the state, but it will remove all entanglement between Alice's ancilla
and the state that she sends. This means that, upon Bob's measurement (he acts like a decohering environment)
we are now back in the situation where Alice sends the state $\ket{C_{a,b}}$ with some probability.

Thus we now show that for a certain distribution $\mu$ over SO(3)
there exists a classical scheme using three-dimensional vectors
that gives rise to an asymptotically secure bit commitment.

We start with an example inspired by \cite{WNI:bc} that
demonstrates that some security gain is easily constructed.
Ref. \cite{WNI:bc} introduces a classical channel that inputs $i\in
\{0,1,2,3\}$ and outputs $i$ with probability $1/2$ and
$i+1\pmod{4}$ with probability $1/2$.  Define codewords
$C_{a,b}=2a+b$ for $a,b\in\{0,1\}$.  Bob accepts if Alice reveals
a codeword that has nonzero (i.e. $1/2$) probability of giving
rise to what he received from the channel.  The protocol is
perfectly sound and concealing, but not binding; Alice can cheat
with probability $1/2$. This security should be compared to the
standard quantum or classical case where being perfectly sound and
perfect concealing implies that Alice can cheat with probability 1.

One use of this channel can be simulated by a frame shift that has
probability $1/2$ of being the identity and probability $1/2$ of
being a rotation about the $z$-axis by $\pi/2$.  The codewords are
$\{\hat{x},\hat{y},-\hat{x},-\hat{y}\}$.

A similar, perhaps more natural, possibility is that the frame shift is a rotation about
the $z$-axis by an angle that is uniformly distributed between $0$
and $\pi$.  Using the same codewords, the protocol is still
perfectly sound and concealing, but Alice can now interpolate
between committing to zero or one by sending the vector
$\cos(\alpha\pi/2)\hat{x} + \sin(\alpha\pi/2)\hat{y}$ for $0\leq
\alpha\leq 1$.  If she then reveals zero, Bob will accept with
probability $1-\alpha/2$, and if she reveals one, Bob will accept
with probability $(1+\alpha)/2$. For Alice, passive cheating is
defined as committing honestly to a bit and then attempting to
reveal something different; here her success probability is still
$1/2$.  However, if Alice chooses $\alpha=1/2$, then she can
convince Bob of either bit with probability $3/4$.

These examples show that misaligned reference frames can offer small
advantages in security over noiseless quantum communication.  But can
we do better?  In the next section, we show that in fact any number of
bits can be committed to any desired level of security for some
distribution $\mu$.

\subsection*{An Asymptotically Secure Scheme}

For our scenario we first assume that Bob can measure the coordinates of a
three-dimensional vector that he receives with infinite precision.
We then argue that the security still holds in the case of finite
precision.

The idea behind the scheme is two-fold. We show that a
single use of a channel can give rise to an (asymptotically)
perfectly secure commitment when the channel acts on vectors in an
arbitrarily high-dimensional space. However, spatial reference
frames are only three-dimensional objects which seems to suggest
that only a partially secure bit commitment may be achievable. Our
scheme overcomes this apparent problem by parametrizing
three-dimensional vectors by coordinates of a $d$-dimensional
lattice.

Alice sends a vector $\vec{v}\in\R^3$
to Bob. In Bob's spatial frame of reference this vector looks like $R \vec{v}$ for
some random rotation $R$ taken from the probability distribution $\mu$.
The distribution $\mu$ of $R$ is the following. Let there be a set of angles
$\theta\in\{\theta_1,\theta_2,\dots,\theta_d\}$ which are {\em not linearly related},
meaning that for all integers (positive or negative)
$n_1,n_2,\dots,n_d$, $\sum_{i=1}^d n_i\theta_i=0 \mod \pi$ if and only if
$n_1=n_2=\dots=n_d=0$. One of these angles $\theta$ is picked uniformly at random. Then $R$ is given by
\be
R= \left\{\begin{array}{ll}
\left(\begin{array}{ccc}\cos\theta& \sin\theta & 0\\ -\sin\theta& \cos\theta & 0 \\ 0 & 0 & 1\end{array}\right) & \mbox{with prob. }1/2, \\
 & \\
\left(\begin{array}{ccc}\cos2\theta& \sin2\theta & 0\\ -\sin2\theta& \cos2\theta & 0 \\ 0 & 0 & 1\end{array}\right)& \mbox{with prob. }1/2.
\end{array}\right.
\ee

Let us describe our bit commitment protocol with this distribution $\mu$, known to Alice and Bob.
Let $L$ be a positive integer.
\begin{itemize}
\item {\sf Commit}. To commit to $b\in\{0,1\}$, Alice chooses
${\bf a}=(a_1,\dots,a_d)\in\{0,1,\dots,L-1\}^d$ uniformly at random, but
conditioned on $\sum_{i=1}^d a_i = b \mod 2$. Let
$\alpha({\bf a}) = \sum_{i=1}^d a_i\theta_i$. Alice sends the vector
$\vec{v}=\vec{v}({\bf a})=(\cos\alpha({\bf a}),\sin\alpha({\bf a}),0)$. Bob receives this vector as $R\vec{v}({\bf a})=\vec{v}({\bf a}')$ with
rotated angle $\alpha'=\sum_i a_i' \theta_i$. He determines the $d$-dimensional
lattice vector ${\bf a}'$. If he cannot find such a vector, he aborts the protocol.
\item {\sf Reveal}. In the reveal phase, Alice simply sends the classical bits
$(b,{\bf a})$ to Bob. Bob accepts when $b = \sum_{i}a_i\mod 2$ and all coordinates of
${\bf a}-{\bf a}'$ are 0 except one for which $a'_i-a_i$ is either 1 or 2. Otherwise
he aborts.
\end{itemize}

Let us now show that this protocol has the desired security properties. \\
{\em Soundness}. We assume that both parties are honest. To understand a given realization of
this protocol, let $j$ be such
that the randomly chosen $\theta$ equals $\theta_j$. In that case the coordinates $a_i'$ of ${\bf a}'$ are
\be
a_i'=\left\{\begin{array}{ll}a_i+\delta_{ij} & \mbox{with prob. }1/2, \\ a_i+ 2 \delta_{ij}& \mbox{with prob. }1/2.\end{array}\right.
\ee
Due to the linear independence of the $\theta$-angles Bob can perfectly compute ${\bf a}'\in
\{0,1,\dots,L+1\}^d$ from $\vec{v}({\bf a}')$. Bob will accept Alice's message in the reveal phase since this vector ${\bf a}'$ is within
distance 2 of the original ${\bf a}$ and the noise acts on only one of the coordinates.

{\em Concealing}. A bit commitment protocol is called
$\epsilon$-concealing when for two different messages $b=0$ and
$b=1$ the distributions over random variables as viewed by Bob are
$\epsilon$-close with respect to their variation distance. In our
case Bob learns ${\bf a}'$, thus we consider the distance
$\epsilon=\sum_{{\bf a}'}|\Pr{{\bf a}'\mid b=0}-\Pr{{\bf a}'\mid
b=1}|$. For an ${\bf a}'$ with all coordinates at least 2 and at
most $L$ we have $\Pr{{\bf a}'\mid b=0}=\Pr{{\bf a}'\mid b=1}$.
All other ${\bf a'}$s are `boundary' cases, which we denote as
${\bf a}' \in {\cal B}$, for which these two conditional
probabilities can be different. We upper bound this boundary term
as
\bea \sum_{{\bf a}' \in {\cal B}} |\Pr{{\bf a'}|b=0}-\Pr{{\bf
a'}|b=1}| \leq \sum_{{\bf a}' \in {\cal B}} \Pr{{\bf a'}}\lesssim \nonumber \\
1-\left(\frac{L-1}{L+2}\right)^d,
\eea
which can be made arbitrarily small for large enough $L$ for any fixed $d$.\\
{\em Binding}. Consider Alice's cheating strategies. She could have sent a different vector, say $\vec{w}(\beta)$.
In case $\vec{w}$ is not in the x-y plane or if $\beta \neq \sum_i b_i \theta_i$ for some integers $b_i \in \{0,\ldots,L+1\}$
Bob simply aborts. Alice could try to cheat by revealing an ${\bf a}^*$ and $b^* \neq b$ that pass Bob's
test. Alice's best option is to choose ${\bf a}^*$ that is the same as ${\bf a}$ except, say,
the $k$th coordinate, which is $a^*_k=a_k+1$. This implies that that the parity $b^*$ of ${\bf a}^*$ is
opposite to $b$. With probability $1/d$ the noise acts on the $k$th coordinate and so
$({\bf a}^*,b^*)$ passes Bob's test. The protocol is $1/d$-binding. \\

\indent In reality we should assume that Bob can only determine $R\vec{v}$ with finite precision,
which means that Bob finds some vector $\vec{w}$ at Euclidean distance $\leq \eps$ from $R\vec{v}$. If
$\eps$ is small enough, we can ensure that for all ${\bf x},{\bf y}\in
\{0,\dots,L+1\}^{d}$, $||\vec{v}({\bf x})-\vec{v}({\bf y})||>2\eps$, so Bob can still determine
${\bf a}'$ from $\vec{w}\approx R\vec{v}$, if Alice behaves honestly.

But what if Alice cheats and sends some arbitrary vector $\vec{w}$ to Bob? Notice, however,
that this strategy could only work if $||R\vec{w} - \vec{v}({\bf a}')||\leq
\eps$, which happens if and only if $||\vec{w} - \vec{v}({\bf a})||\leq \eps$. In
particular, ${\bf a}$ is the {\em only} $d$-dimensional vector such
that $\vec{v}({\bf a})$ is $\eps$-close to $\vec{w}$. Thus if Alice later
reveals ${\bf a}^*\neq {\bf a}$, her cheating still succeeds only with
probability $\leq 1/d$.

{\em Remark:} By increasing $d$ and running the above protocol in
parallel several times, it is also possible to commit more than one
bit.

It remains an open problem to get a more complete overview of the (im)possibility of bit
commitment for general distributions $\mu$.
In particular it would be interesting to find realistic noise
models for, say, polarized photons, that would allow for secure or approximately secure bit commitment.

RO and BMT would like to acknowledge support by the NSA and the ARDA through ARO contract number
W911NF-04-C-0098. AWH thanks the IBM Watson quantum information group for their
hospitality while doing this work.

% AH acknowledgements?

\bibliographystyle{hunsrt}
\bibliography{refs}

\begin{thebibliography}{10}

\bibitem{BB:84}
C.~H. Bennett and G.~Brassard.
\newblock Quantum cryptography: Public key distribution and coin tossing.
\newblock In {\em Proceedings of the IEEE International Conference on
  Computers, Systems and Signal Processing}, pages 175--179, 1984.

\bibitem{mayers:qbcsecurity}
D.~Mayers.
\newblock Unconditionally secure quantum bit commitment is impossible.
\newblock {\em Phys. Rev. Lett.}, 78:3414--3417, 1997.

\bibitem{LC:qbcsecurity}
H.-K. Lo and H.~Chau.
\newblock Is quantum bit commitment really possible?
\newblock {\em Phys. Rev. Lett.}, 78:3410--13, 1997.

\bibitem{LC:qbcsecurity2}
H.-K. Lo and H.~Chau.
\newblock Why quantum bit commitment and ideal quantum coin tossing are
  impossible.
\newblock {\em Physica D}, 120:177, 1998.

\bibitem{KMP:ss}
A.~Kitaev, D.~Mayers, and J.Preskill.
\newblock Superselection rules and quantum protocols.
\newblock {\em Phys. Rev. A}, 69:052326, 2004,
  \url{http://arxiv.org/abs/quant-ph/0310088}.

\bibitem{blum:bc}
Manuel Blum.
\newblock Coin flipping by telephone a protocol for solving impossible
  problems.
\newblock {\em SIGACT News}, 15(1):23--27, 1983.

\bibitem{naor:bc}
M.~Naor.
\newblock Bit commitment using pseudorandomness.
\newblock {\em Journal of Cryptology}, 4(2):151--158, 1991.

\bibitem{WNI:bc}
A.~Winter, A.C.A. Nascimento, and H.~Imai.
\newblock Commitment capacity of discrete memoryless channels.
\newblock In {\em IMA Int. Conf.}, pages 35--51. Springer, 2003,
  \url{http://arxiv.org/abs/cs.CR/0304014}.

\bibitem{PT:rel}
A.~Peres and D.~Terno.
\newblock Quantum information and relativity theory.
\newblock {\em Rev. of Mod. Phys.}, 76:93, 2004,
  \url{http://arxiv.org/abs/quant-ph/0212023}.

\bibitem{PS:dir}
A.~Peres and P.~Scudo.
\newblock Entangled quantum states as direction indicators.
\newblock {\em Phys. Rev. Lett.}, 86:4160, 2001,
  \url{http://arxiv.org/abs/quant-ph/0010085}.

\bibitem{RG:cc}
T.~Rudolph and L.~Grover.
\newblock On the communication complexity of establishing a shared reference
  frame.
\newblock {\em Phys. Rev. Lett.}, 91:217905, 2003,
  \url{http://arxiv.org/abs/quant-ph/0306017}.

\bibitem{CPS:refframe}
G.~Chiribella, G.M. D'Ariano, P.~Perinotti, and M.F. Sacchi.
\newblock Efficient use of quantum resources for the transmission of a
  reference frame.
\newblock {\em Phys. Rev. Lett.}, 93:180503, 2004,
  \url{http://arxiv.org/abs/quant-ph/0405095}.

\bibitem{vanenk:refbit}
S.~van Enk.
\newblock Quantifying the resource of sharing a reference frame.
\newblock {\em Phys. Rev. A}, 71:032339, 2004,
  \url{http://arxiv.org/abs/quant-ph/0410083}.

\bibitem{BRS:comm}
S.D. Bartlett, T.~Rudolph, and R.W. Spekkens.
\newblock Classical and quantum communication without a shared reference frame.
\newblock {\em Phys. Rev. Lett.}, 91:027901, 2003,
  \url{http://arxiv.org/abs/quant-ph/0302111}.

\bibitem{BRS:cryp}
S.D. Bartlett, T.~Rudolph, and R.W. Spekkens.
\newblock Decoherence-full subsystems and the cryptographic power of a private
  shared reference frame.
\newblock {\em Phys. Rev. A}, 70:032307, 2004,
  \url{http://arxiv.org/abs/quant-ph/0403161}.

\end{thebibliography}

\end{document}